\documentclass[useAMS,usenatbib]{mn2e}

\usepackage{graphicx}
\usepackage{epsfig,floatflt}

\title[Binarity among Magellanic Cepheids]
{Binarity among the Cepheids in the Magellanic Clouds}
\author[L. Szabados and D. Neh\'ez]{L\'aszl\'o Szabados$^{1}$\thanks{E-mail:
szabados@konkoly.hu} and D\'ora Neh\'ez$^{2}$
\\
$^{1}$Konkoly Observatory,
Research Centre for Astronomy and Earth Sciences,
Hungarian Academy of Sciences,\\
H-1121 Budapest, Konkoly Thege Mikl\'os \'ut 15-17, Hungary\\
$^{2}$Department of Astronomy, Lor\'and E\"otv\"os University,
H-1117 Budapest, P\'azm\'any P. s\'et\'any, Hungary
}

\begin{document}

\date{Accepted Received ; in original form }

\pagerange{\pageref{firstpage}--\pageref{lastpage}} \pubyear{2012}

\maketitle

\label{firstpage}

\begin{abstract}
Spectroscopic binarity of the Cepheid variable HV914 in 
the Large Magellanic Cloud is pointed out from the published radial 
velocity observational data. The list of known binaries among 
Cepheid type variable stars in the Magellanic Clouds is
published in tabular form. The census indicates a serious deficiency 
of Cepheids with known companions as compared with their 
Galactic counterparts, whose implications are also discussed. 
A particular amplitude ratio ($A_{V_{\rm rad}}/A_B$) of
individual Magellanic Cepheids is studied in order to select
promising candidates of spectroscopic binaries worthy of thorough
radial velocity studies.
\end{abstract}


\section{Introduction}
\label{intro}

Extragalactic Cepheid variable stars are key objects in astronomy
because they are {\em standard candles} (primary distance indicators)
owing to the correlation between their pulsation period, $P$, and 
luminosity, $L$ -- the well known $P$-$L$ relationship. 

The Magellanic Clouds have been pivotal objects in the permanent 
recalibration of the $P$-$L$ relationship because the slope of 
this relationship is usually based on Cepheids in these two 
nearby extragalactic systems, while several carefully selected Galactic 
Cepheids serve as calibrators of the zero-point of the relationship.

When attempting to improve the calibration, one of the main goals is 
to decrease the spread of the points around the ridge-line fit. 
The dispersion of the relationship is caused by various factors
\citep[recently summarized by][]{SzK12}.
One of these factors is the presence of a companion to the Cepheid.

The frequency of occurrence of binaries among Galactis Cepheids 
exceeds 50 per cent \citep{Sz03}.
There is, however, a strong observational selection effect
that hinders the discovery of binarity, as pointed out by \citet{Sz03}.
Owing to their distances, Cepheids in the Magellanic Clouds appear 
much fainter than their Galactic counterparts. One of the
consequences of this fact is the extremely low number of Cepheids
known to be members in binary systems in the Magellanic Clouds (see 
Sect.~\ref{census}).

Binarity of Cepheids can be revealed by spectroscopic, photometric, 
or astrometric methods as summarized by \citet{SzK12}. 
In the case of Cepheids in the Magellanic Clouds, the
spectroscopic method is the most efficient one in revealing
binarity. The spectroscopic signs that hint at the presence of a 
companion include:\\
- a secondary star hotter than the Cepheid component results in 
an excess flux in the UV region as compared to the normal Cepheid 
spectrum;\\
- variations in the radial velocity due to the orbital motion
superimposed on those of pulsational origin.\\
\noindent Cepheids in the Magellanic Clouds are too faint 
for UV spectroscopy with the current and former equipments (e.g.,
the {\it IUE\/} space mission dedicated for UV spectroscopy).
Optical spectroscopy can be, however, instrumental in increasing 
the number of known spectroscopic binaries among Magellanic 
Cepheids. The orbital effect can be observed if the line of sight 
towards the Cepheid is nearly parallel or subtends a small angle 
to the orbital plane of the spectroscopic binary system. 

In this paper we point out spectroscopic binary nature of  
HV914 in the Large Magellanic Cloud (LMC) 
(Sect.~\ref{HV914}).
Based on a comprehensive search for relevant data and papers
in the literature, we publish a state-of-the-art census of known 
binaries among Magellanic Cepheids (Sect.~\ref{census}) and
select several suspected binaries suggested for dedicated
spectroscopic studies (Sect.~\ref{susp}).

\section{HV914}
\label{HV914}

\begin{figure}
\includegraphics[height=55mm, angle=0]{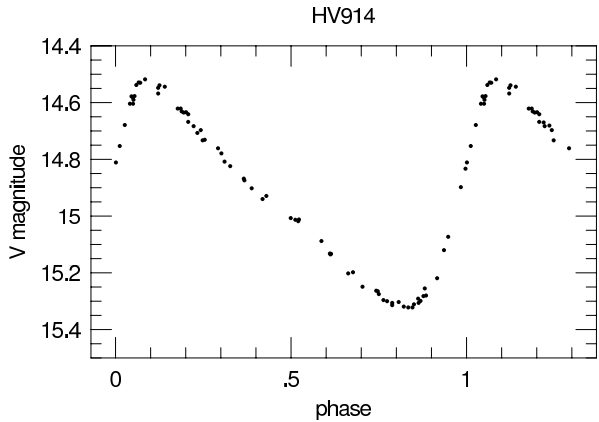}
\caption{OGLE-III photometric $V$ phase curve of HV914 \citep{Setal08a}.
The data have been folded on the period rounded to 6.8784\,d, the
zero phase set arbitrarily at JD\,2\,400\,000}
\label{fig_lc_HV914}
\end{figure}

The brightness variability of this Cepheid located in the 
LMC was discovered by \citet{L08} based on Harvard
photographic plates. The Cepheid nature of variability 
and the first value of the pulsation period (6.8795\,d) 
was published by \citet{PG71}. Further observations of
this Cepheid have not been performed until the commencement
of the microlensing projects. The analysis of the extensive 
OGLE-III photometric data set (obtained in $V$ and $I$ bands) 
resulted in an accurate period of 6.8783932\,d \citep{Setal08a}. 
The phase curve in $V$ band is plotted in Fig.~\ref{fig_lc_HV914}.

Recently, HV914 was included in a spectroscopic observational 
project of Magellanic Cepheids \citep{Setal11}. The impressive 
radial velocity phase curves published 
by \citet{Setal11} densely cover the whole pulsation cycle for 
all target Cepheids and the individual data scatter around 
the normal curve by less than 0.1 km\,s$^{-1}$ for the 22 program 
stars except HV914 \citep[see Fig.~2 in][]{Setal11}. 

In the case of classical Cepheids, such excessive dispersion 
of the radial velocity phase curve can be caused by:\\
- using an inaccurate period when plotting the phase curve;\\
- an additional excited mode;\\
- spectroscopic binarity.

The value of the pulsation period is accurately known
from the precise OGLE-III photometric data covering the years
1997-2008. Because the radial velocity data by \citet{Setal11}
are based on observations performed during the years 2000-2003,
the same period can be assumed as obtained from the OGLE-III data:
6.8783932\,d. In fact, \citet{Setal11} plotted their radial
velocity data for HV914 using the period of 6.878425\,d which is
not identical with the previous value but this slight difference
(of 0.000032\,d) cannot be responsible for the dispersion 
exceeding 5 km\,s$^{-1}$ seen in the radial velocity phase curve.

The high-quality light curve (Fig.~\ref{fig_lc_HV914}) testifies 
that HV914 is a single-mode Cepheid. An indirect additional evidence 
in favour of the single-mode pulsation is the fact, that secondary
periods are only excited in Cepheids with fundamental period shorter
than 4 days in the LMC \citep{Setal10a}.

\begin{figure}
\includegraphics[height=55mm, angle=0]{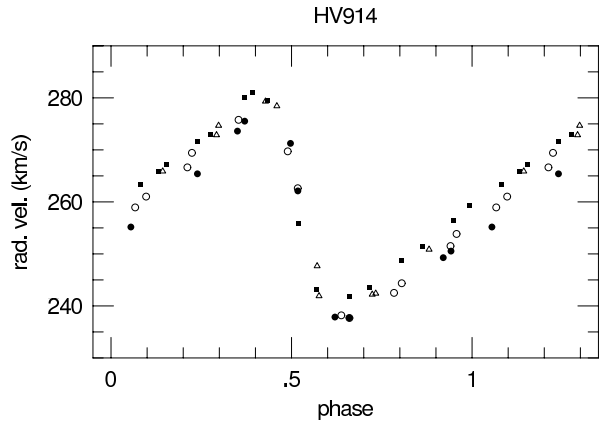}
\caption{Radial velocity phase curve of HV914 based on the data
published by \citet{Setal11}. Data representing the 1999/2000 observing
season are marked with filled squares; 2000/2001 data: open triangles;
2001/2002 data: open circles; 2002/2003 data: filled circles}
\label{fig_vrad_HV914}
\end{figure}

\begin{figure}
\includegraphics[height=55mm, angle=0]{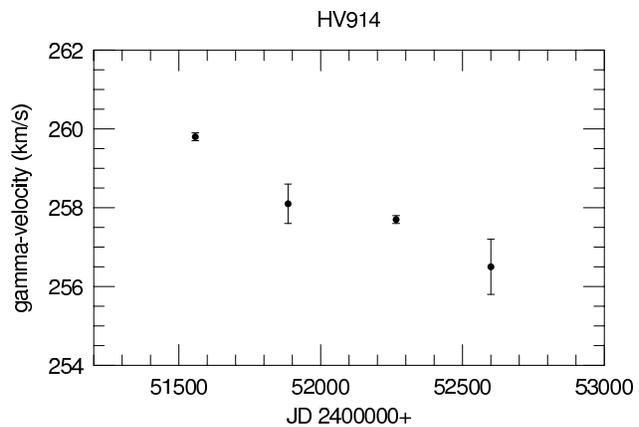}
\caption{Temporal drift in the $\gamma$-velocity of HV914}
\label{v-gamma-HV914}
\end{figure}

If HV914 has a physical companion orbiting the Cepheid component,
the orbital period cannot be shorter than 1-2 years, because 
Cepheids are supergiant stars. In this case a seasonal shift is 
expected in the radial velocity phase curve whose amount depends on 
the orientation of the orbital plane with respect 
to the line of sight. In fact, such seasonal shifts are noticeable 
in Fig.~\ref{fig_vrad_HV914}. The thousand-day observational
interval was not sufficiently long to cover a complete orbital cycle.
But even in this case, the monotonic shift of the centre-of-mass
radial velocity ($\gamma$-velocity) 
plotted in Fig.~\ref{v-gamma-HV914}
refers to the orbital motion, i.e., to spectroscopic binarity of HV914.

\begin{table*}
\caption{Known binaries among Cepheids in the Large Magellanic Cloud}
\begin{tabular}{l@{\hskip2mm}c@{\hskip2mm}l@{\hskip2mm}l}
\hline
\noalign{\vskip 0.1mm}
Cepheid& $P_{\rm puls}$ & Remarks & Reference\\
& (day) & Type (orbital period) [cross-ID] &\\
\noalign{\vskip 0.1mm}
\hline
\noalign{\vskip 0.1mm}
HV883 &              133.9   &    SB &\citet{I94}\\
HV914 &               6.8784 &   SB [OGLE-SMC-CEP-1249]&present paper\\
HV12202  &           3.10112 &    SB &\citet{Wetal91}, \citet{Setal05}\\
HV12204  &           3.43876 &    SB &\citet{Wetal91}, \citet{Setal05}\\
OGLE-LMC-CEP-0227 &  3.79709 &    EB (309.673 d) &\citet{Petal10}\\
OGLE-LMC-CEP-1812 &  1.31290 &    EB (552.0 d) &\citet{Setal08a}\\
OGLE-LMC-CEP-2532 &  2.03536 &    EB (800.5 d) [MACHO 81.8997.87]
&\citet{Setal08a}, \citet{Aetal02}\\
OGLE-LMC-CEP-1718  & 1.96366+2.48094 & Double Cepheid + EB (825.6 d)&\citet{Setal08a}\\
OGLE-LMC-T2CEP-021 &      9.7596   &       EB (174.83 d)&\citet{Setal08b}\\
OGLE-LMC-T2CEP-023 &      5.2348   &       EB (88.51 d) &\citet{Setal08b}\\
OGLE-LMC-T2CEP-052 &      4.66795  &       EB (123.81 d)&\citet{Setal08b}\\
OGLE-LMC-T2CEP-077 &      1.213802 &       EB (34.621 d) &\citet{Setal08b}\\
OGLE-LMC-T2CEP-084 &      1.770840 &       EB (52.355 d)&\citet{Setal08b}\\
OGLE-LMC-T2CEP-093 &      17.5930  &       EB (419.9 d) [MACHO
78.6338.24]&\citet{Setal08b}, \citet{Aetal02}\\
OGLE-LMC-T2CEP-098 &      4.97374  &       EB (397.2 d) [MACHO
6.6454.5]&\citet{Setal08b}, \citet{Aetal02}\\
MACHO$\star$04:59:17.5 $-$69:14:18 & 2.1010+3.0805 & Double Cepheid
&\citet{Aetal95}\\
MACHO$\star$05:04:02.3 $-$68:21:32 & 2.7509+4.5630 & Double Cepheid
&\citet{Aetal95}\\
\hline
\end{tabular}
\label{tab_LMC}
\end{table*}

\begin{table*}
\caption{Known binaries among Cepheids in the Small Magellanic Cloud}
\begin{tabular}{l@{\hskip2mm}c@{\hskip2mm}l@{\hskip2mm}l}
\hline
\noalign{\vskip 0.1mm}
Cepheid& $P_{\rm puls}$ & Remarks & Reference\\
& (day) & Type (orbital period) [cross-ID] &\\
\noalign{\vskip 0.1mm}
\hline
\noalign{\vskip 0.1mm}
HV837               & 42.6789  &     SB&\citet{I94}\\
HV11157             & 69.0       &   SB&\citet{I94}\\
OGLE-SMC-CEP-0411   & 1.10098    &   EB (43.498 d)&\citet{Setal10a}\\
OGLE-SMC-CEP-1526   & 1.29023 + 1.80431 & Double Cepheid [SMC$\_$SC5$\_$208044]&\citet{Setal10a}\\
OGLE-SMC-CEP-1996   & 2.31795     &  EB (95.594 d)&\citet{Setal10a}\\
OGLE-SMC-CEP-2699   & 2.11748 + 2.56230 & Double Cepheid&\citet{Setal10a}\\
OGLE-SMC-CEP-2893   & 1.13586 + 1.32155 & Double Cepheid&\citet{Setal10a}\\
OGLE-SMC-CEP-3115   & 1.15979 + 1.25194 & Double Cepheid&\citet{Setal10a}\\
OGLE-SMC-CEP-3674   & 1.82776 + 2.89605 & Double Cepheid&\citet{Setal10a}\\
OGLE-SMC-T2CEP-007  & 30.9606    &   ELL (392.93)&\citet{Setal10b}\\
OGLE-SMC-T2CEP-010  & 17.4807    &   ELL (198.18 d)&\citet{Setal10b}\\
OGLE-SMC-T2CEP-023  & 17.6753    &   EB (156.884 d)&\citet{Setal10b}\\
OGLE-SMC-T2CEP-025  & 14.17089   &   ELL (174.87)&\citet{Setal10b}\\
OGLE-SMC-T2CEP-028  & 15.2643    &   EB (141.835 d)&\citet{Setal10b}\\
OGLE-SMC-T2CEP-029  & 33.6765    &   EB (608.6 d)&\citet{Setal10b}\\
\noalign{\vskip 0.1mm}
\hline
\end{tabular}
\label{tab_SMC}
\end{table*}

\section{Known binaries among the Magellanic Cepheids}
\label{census}

There are 3361 known classical Cepheids in the LMC \citep{Setal08a} 
and 4630 classical Cepheids are known in the SMC \citep{Setal10a}.
Tables~\ref{tab_LMC}-\ref{tab_SMC} list all non-solitary Cepheids 
(i.e., classical and Type~II Cepheids, as well) known in the LMC 
and SMC, respectively. In view of the high incidence of binaries 
among Galactic classical Cepheids, there is a serious deficiency 
of known binaries among Magellanic Cepheids. 

In the LMC, four Cepheids are members in spectroscopic 
binaries, 11 Cepheids show eclipsing variability superimposed
on the pulsational variations. Furthermore, there are 3 double 
Cepheids which probably form physical pairs consisting of two 
Cepheid components. Such pairs are similar to the Galactic binary 
Cepheid CEa~Cas + CEb~Cas. This archetype of Cepheid pairs
is described by \citet{Oetal88}. 
Interestingly, one of the double Cepheids in the LMC shows 
eclipsing light variations, as well, i.e., this system consists of 
three components. Triple systems frequently occur among Galactic 
classical Cepheids, see the on-line data base of binary Cepheids:
http://www.konkoly.hu/CEP/intro.html.

As to the SMC, there are 4 known spectroscopic binaries and 5 
eclipsing binaries involving a Cepheid component, as well as 
5 double Cepheids. In addition, 3 Cepheids show variability typical 
of ellipsoidal binaries superimposed on the pulsational variations.

The frequency of known binaries among Magellanic Cepheids
is at least two orders of magnitudes lower than the corresponding 
frequency of binaries among Galactic classical Cepheids. 
This must be an observational selection effect.
Therefore, discovery of a plethora of new binary systems among 
Magellanic Cepheids is expected from a suitable observational material.

\begin{table*}
\caption{Amplitude ratios for Cepheids in the Large Magellanic Cloud}
\begin{tabular}{lrcccccccl}
\hline
\noalign{\vskip 0.1mm}
Cepheid& Period & $A_B$ & $A_V$ & $A_{V_{\rm rad}}$ & $q$ & $A_V/A_B$ &$q'$ &
Mode &Sources\\
 & (d)& (m)& (m)& (km/s)& &&&&\\
\noalign{\vskip 0.1mm}
\hline
\noalign{\vskip 0.1mm}
HV873 &34.46   &  -  &1.328 &49.58 &  -   &  -   &24.64 &F&6, 10, 13\\
HV867 &22.72   &  -  &1.179 &61.34 &  -   &  -   &34.34 &F&10, 13\\
HV877 &45.29   &0.999&0.607 &30.47 &30.50 &0.608 &33.13 &F&6, 7, 13\\
HV878 &23.30   &1.969&1.187 &60.34 &30.64 &0.603 &33.55 &F&6, 10, 13\\
HV879 &36.82   &1.896&1.197 &54.29 &28.63 &0.631 &29.93 &F&4, 7, 8\\
HV881 &35.74   &1.677&1.285 &50.96 &30.39 &0.766 &26.17 &F&6, 10, 13\\
HV883 &132.26  &1.671& -    &62.86 &37.62 &  -   &  -   &F&2, 3, 5, 7, 8, 10\\
HV899 &31.32   &1.963&1.221 &58.91 &30.01 &0.622 &31.84 &F&4, 6, 7, 8\\
HV900 &48.02   &  -  &0.940 &46.58 &  -   &  -   &32.71 &F&5, 6, 7, 8, 13\\
HV909 &37.54   &1.702& -    &44.21 &25.98 &  -   &  -   &F&4, 6, 7, 8\\
HV914 &6.88    &  -  &1.006 &43.52 &  -   &  -   &28.55 &F&13\\
HV1005 &18.72  &1.796&1.111 &62.44 &34.77 &0.619 &37.09 &F&7, 10, 13\\
HV1006 &14.22  &  -  &1.241 &54.35 &  -   &  -   &28.90 &F&10, 13\\
HV1023 &26.56  &1.508&1.017 &52.04 &34.51 &0.674 &33.77 &F&7, 13\\
HV2257 &39.37  &1.873&1.179 &50.79 &27.12 &0.629 &28.43 &F&4, 6, 7, 8, 10\\
HV2282 &14.68  &  -  &1.040 &53.42 &  -   &  -   &33.90 &F&10, 13\\
HV2338 &42.20  &1.916& -    &51.17 &26.71 &  -   &  -   &F&4, 6, 7, 8\\
HV2369 &48.36  &1.830&1.233 &47.10 &25.74 &0.674 &25.21 &F&2, 6, 7, 13\\
HV2405 &6.29   &  -  &0.478 &25.25 &  -   &  -   &34.86 &F&13\\
HV2447 &118.20 &0.913& -    &27.74 &30.38 &  -   &  -   &F&2, 5, 6, 7, 8\\
HV2527 &12.95  &1.585&1.030 &53.45 &33.72 &0.560 &34.25 &F&7, 10, 13\\
HV2538 &13.87  &  -  &0.544 &32.01 &  -   &  -   &38.84 &F&13\\
HV2549 &16.22  &1.546&1.135 &40.76 &26.36 &0.734 &23.70 &F&7, 10, 13\\
HV2694 &6.94   &  -  &0.937 &42.81 &  -   &  -   &30.15 &F&11\\
HV2827 &78.78  &0.910& -    &26.57 &29.20 &  -   &  -   &F&4, 7, 8\\
HV2883 &108.93 &1.974& -    &49.07 &24.86 &  -   &  -   &F&2, 5, 6, 7, 8\\
HV5497 &99.45  &0.838& -    &25.58 &30.53 &  -   &  -   &F&2, 5, 6, 7, 8\\
HV5655 &14.21  &  -  &0.956 &54.37 &  -   &  -   &37.54 &F&7, 13\\
HV6093 &4.78   &  -  &0.834 &41.24 &  -   &  -   &32.64 &F&13\\
HV12197 &3.14  &0.835& -    &35.13 &42.07 &  -   &  -   &1OT&9, 12, 14, 15\\
HV12198 &3.52  &0.927& -    &42.59 &44.87 &  -   &  -   &1OT&5, 11, 12, 14, 15\\
HV12199 &2.64  &1.021& -    &44.92 &44.00 &  -   &  -   &1OT&12, 14, 15\\
HV12202 &3.10  &0.682& -    &53.80 &78.89 &  -   &  -   &1OT&12, 14, 15\\
HV12203 &2.95  &0.950& -    &38.91 &40.96 &  -   &  -   &1OT&12, 14, 15\\
HV12204 &3.44  &1.002& -    &52.88 &52.77 &  -   &  -   &1OT&12, 14, 15\\
HV12452 &8.75  &  -  &0.852 &37.91 &  -   &  -   &29.37 &F&13\\
HV12505 &14.39 &  -  &0.911 &52.31 &  -   &  -   &37.90 &F&10, 13\\
HV12717 &8.84  &  -  &0.823 &37.47 &  -   &  -   &30.05 &F&10, 13\\
HV12815 &26.12 &1.813& -    &59.85 &33.01 &  -   &  -   &F&1, 7, 8, 10\\
HV12816 &9.10  &1.041& -    &29.15 &28.00 &  -   &  -   &F&1, 7, 8\\
NGC1866-V4 &3.34 &0.405& -  &17.46 &43.11 &  -   &  -   &F&12, 15\\
NGC1866-We8 &3.04 &0.879& - &34.90 &44.23 &  -   &  -   &F&15\\
U1      &22.54 &1.853&1.116 &67.29 &36.31 &0.602 &39.80 &F&10, 13\\
\noalign {\vskip 0.1mm}
\hline
\end{tabular}

1 - \citet{Cetal86};
2 - \citet{Fetal85};
3 - \citet{I94}; 
4 - \citet{Ietal85};
5 - \citet{Ietal89};
6 - \citet{M75};
7 - \citet{MW79};
8 - \citet{Metal98};
9 - \citet{Metal12};
10 - \citet{Setal02};
11 - \citet{Setal04};
12 - \citet{Setal05};
13 - \citet{Setal11};
14 - \citet{W87};
15 - \citet{Wetal91}
\label{tab_LMCampl}
\end{table*}

\begin{table*}
\caption{Amplitude ratios for Cepheids in the Small Magellanic Cloud}
\begin{tabular}{lrcccccccl}
\hline
\noalign{\vskip 0.1mm}
Cepheid& Period & $A_B$ & $A_V$ & $A_{V_{\rm rad}}$ & $q$ & $A_V/A_B$ &$q'$ &
Mode & Sources\\
 & (d)& (m)& (m)& (km/s)& &&&&\\
\noalign{\vskip 0.1mm}
\hline
\noalign{\vskip 0.1mm}
HV821  &128.0 &0.972 &  -   &32.18 &33.11 & -    & -    &F&1, 3, 5, 6, 7, 8\\
HV822  & 16.7 & -    &1.172 &57.87 &  -   & -    &32.58 &F&6, 9\\
HV824  & 65.9 &1.431 &  -   &43.08 &30.10 & -    & -    &F&1, 5, 7, 8\\
HV829  & 85.3 &1.043 &  -   &34.02 &32.62 & -    & -    &F&1, 3, 5, 6, 7, 8\\
HV834  & 73.6 &1.357 &  -   &51.41 &37.89 & -    & -    &F&1, 3, 5, 6, 7, 8\\
HV837  & 42.7 &1.434 &0.885 &47.87 &33.82 &0.617 &35.70 &F&1, 5, 6, 7, 8\\
HV1328 & 15.8 &1.072 &0.772 &24.29 &22.66 &0.720 &20.77 &F&9\\
HV1333 & 16.3 & -    &0.931 &50.35 &  -   & -    &35.69 &F&1, 9\\
HV1335 & 14.4 &1.264 &0.814 &38.14 &30.17 &0.644 &30.92 &F&1, 9\\
HV1338 &  8.5 &1.293 &0.888 &37.87 &29.29 &0.687 &28.15 &F&1, 2, 8\\
HV1345 & 13.5 &1.184 &0.795 &36.10 &30.49 &0.671 &29.97 &F&9\\
HV1365 & 12.4 &1.049 &0.835 &41.85 &39.90 &0.796 &33.08 &F&1, 2, 7, 8\\
HV11157& 69.2 &0.653 &0.382 &24.09 &36.89 &0.585 &41.62 &F&1, 3, 4, 5, 6, 7, 8\\
\noalign{\vskip 0.1mm}
\hline
\end{tabular}

1 - \citet{CC84};
2 - \citet{Cetal86};
3 - \citet{Fetal85};
4 - \citet{I94}; 
5 - \citet{Ietal89};
6 - \citet{M75};
7 - \citet{MW79};
8 - \citet{Metal98};
9 - \citet{Setal04}
\label{tab_SMCampl}
\end{table*}


\section{Selection of more candidate binaries}
\label{susp}

To facilitate the selection of promising binary
candidates among Magellanic Cepheids, we applied the amplitude 
ratio criterion \citep[see][]{KSz09}. This method is based on 
the fact that a companion star decreases the observable
photometric amplitude while the amplitude of radial
velocity variations ($A_{V_{\rm rad}}$) remains unaffected, 
or an unrevealed orbital motion superimposed on the $V_{\rm rad}$
variations of pulsational origin even increases the observable
peak-to-peak amplitude. A thorough study of more than 200
Galactic Cepheids \citep{KSz09} has shown that the average value
of the amplitude ratio $A_{V_{\rm rad}}/A_B$ (where $A_B$ is
the peak-to-peak photometric amplitude in the Johnson $B$ band),
denoted by $q$ is 32.79 for fundamental mode Cepheids without 
known companions and 35.23 for first overtone Cepheids, also
without companions. Significantly larger $q$ values may hint
at the presence of companion(s).

To determine the $q$ value of Magellanic Cepheids, we 
collected the $B$ and $V$ photometric data and the
spectroscopic radial velocity observational data published
in the literature or accessible from public online data bases.
We redetermined the pulsation period using the program package
{\sc mufran} \citep{K91}. The peak-to-peak amplitudes were determined
from the phase curve folded with this period and the Fourier fit 
taking into account the shape of the phase curve (at certain
period values the Hertzsprung progression of Cepheids results
in rather sharp features in the phase curves that may affect the
amplitude). The results are summarized in Tables~\ref{tab_LMCampl} 
and~\ref{tab_SMCampl} for the Cepheids in the LMC and SMC,
respectively. The individual columns of both tables contain
the following data:\\
1. - Identification of the Cepheid;\\
2. - Pulsation period (in days) rounded to two decimals;\\
3. - Peak-to-peak amplitude in the Johnson $B$ band, $A_B$;\\
4. - Peak-to-peak amplitude in the Johnson $V$ band, $A_V$;\\
5. - Peak-to-peak radial velocity amplitude,  $A_{V_{\rm rad}}$;\\
6. - The $q$ amplitude ratio;\\
7. - The ratio of amplitudes in $V$ and $B$ bands, $A_V/A_B$;\\
8. - The $q'$ amplitude ratio explained in the text;\\
9. - Pulsation mode (F for fundamental mode, 1OT for first overtone
pulsation;\\
10. - Sources of the observational data listed as footnotes to the
respective tables.

This $q$-method involves the $B$ photometric amplitude
because the overwhelming majority of Cepheid companions are blue 
stars thus the diminution of the amplitude in $B$ is more prominent 
than in the $V$ band. $B$-band photometry for Magellanic Cepheids
is often unavailable because extragalactic Cepheids are
much fainter than their Galactic counterparts. Extensive photometry 
in $V$ band is, however, available in the OGLE-III data base 
(http://ogledb.astrouw.edu.pl/$\sim$ogle/CVS/). The $V$ amplitudes
in Tables~\ref{tab_LMCampl} and~\ref{tab_SMCampl} are listed 
based on the analysis of these data without mentioning this 
source in the last column.

The known photometric amplitude in the $V$ band allowed us to compute 
a `pseudo'-$q$ value based on the fact that there exists a close 
correlation between the $V$ and $B$ photometric
amplitudes of Cepheids: the ratio of $A_V/A_B = 0.685\pm0.004$ for 
Galactic Cepheids without known companions \citep{KSz09}. 
The corresponding amplitude ratio is $A_V/A_B = 0.66\pm0.01$ for
the Magellanic Cepheids involving both Clouds from the data listed
in Tables~\ref{tab_LMCampl} and~\ref{tab_SMCampl}. Here binary
Cepheids have not been excluded causing a slight increase in the 
actual value of the ratio that, however, can be neglected for our 
purpose. The missing $A_B$ value was then replaced by $A_V/0.66$
in calculating the $q$ parameter. The value derived in such a way
is denoted as $q'$, and this was referred to as the `pseudo'-$q$ 
value.

Again, an excessive value of $q'$ can be a sign of a companion. 
Those Cepheids which have both $q$ and $q'$ values show reliability 
of the $q$ parameter, and the large value of either $q$ or $q'$
parameter testifies usefulness of this binarity indicator. The
known spectroscopic binaries HV883, HV12202, HV12204 in the LMC, as
well as HV11157 in the SMC all have much larger $q$ and/or $q'$ 
values than the corresponding average for Cepheids of the given
pulsation mode. Moreover, HV837 is an example for a Cepheid
in a spectroscopic binary that has an almost normal $q$ value.

There are other Cepheids showing excessive $q$ and/or $q'$ values: 
HV1005, HV2538, HV5655, HV12505, NGC1866-V4, NGC1866-We8, and U1
in the LMC, as well as HV1365 and HV834 in the SMC. 
All of them are fundamental mode pulsators. A dedicated radial 
velocity study of these Cepheids can lead to the discovery of 
their spectroscopic binary nature.

Some radial velocity data are available for each of these 
candidate binaries.\\
\noindent - For HV2538, HV5655, and HV12505 the data were
collected during a time interval as short as two weeks in 2005
\citep{Setal11}, thus a second epoch radial velocity series 
can help reveal a variable $\gamma$-velocity.\\
\noindent - The 7 $v_{\rm rad}$ values published for NGC1866-V4 cover 
only a week \citep{Wetal91}.\\
\noindent - There are 53 radial velocity data on HV834 obtained between 
1981 and 1985 \citep{Ietal89} but the phase curve does not indicate any
variability in the $\gamma$-velocity. 

For the other four candidates, there are some pieces of evidence of 
binarity based on the available radial velocity data.\\
\noindent - HV1005 was observed by \citet{Setal11} mostly in 2005, 
and the additional three points obtained in 2007 significantly deviate 
from the phase curve plotted for the 2005 data.\\
\noindent -  Data for U1 were also obtained by \citet{Setal11} in three 
different observing seasons. Unfortunately, different parts of the 
phase curve were covered in each season, so there is no apparent 
dispersion of the data points but the shape of the phase curve is 
untypical of the given pulsation period. However, the strange shape 
of the phase curve can be reconciled by some vertical shift of the 
data obtained in different years, i.e., by ``adjusting'' the 
$\gamma$-velocity.\\
\noindent - The available radial velocity data on HV1365 were obtained 
in three observing seasons \citep{Cetal86}. An annual shift in the 
$\gamma$-velocity is suspected that has to be confirmed by new and
more accurate radial velocity data.\\
\noindent - The clearest (though still weak) spectroscopic evidence 
of binarity exists for NGC1866-We8, whose radial velocity phase plot 
(based on 6 data) contains a deviating point (from 1989) with respect 
to the 5 observational data obtained in 1988 
\citep[see Fig.~30 in][]{Wetal91}.

Revealing binarity of individual Cepheids in the Magellanic Clouds
is important because the calibration of the $P$-$L$ relationship 
relies on the slope determined from Magellanic Cepheids. Companions 
to Cepheids increase the dispersion of the resulting $P$-$L$ 
relationship. Moreover, shorter period Cepheids are more affected
by the adverse photometric effects because of smaller luminosity
difference between the Cepheid and its companion. This may even
cause a period dependent slope of the $P$-$L$ relationship.

To avoid such distorsions, Cepheids with known companions have 
to be excluded from the calibration of the $P$-$L$ relationship. 
If, however, physical properties of the companions can be determined, 
the $P$-$L$ relationship can be calibrated based on the binary 
Cepheids themselves \citep[see e.g.,][]{E92}.

\section{Conclusion}
\label{conclusion}
Spectroscopic binarity of HV914, a Cepheid in the Large
Magellanic Cloud has been revealed from existing radial velocity 
data. Further accurate data are necessary to determine the 
orbital elements. 

The census of Cepheids known to belong to binary systems is also 
studied for both Magellanic Clouds. 
The serious deficiency of binary member Cepheids (with respect 
to their Galactic counterparts) is explained by the lack of dedicated 
observational studies. Radial velocity observations of a large 
number of Cepheids in the Magellanic Clouds, especially those having
high values of the $q$ and/or $q'$ amplitude ratios (listed in this
paper) are desirable.

\section*{Acknowledgments}
Financial support from the ESA PECS project C98090 and
ESTEC Contract No.4000106398/12/NL/KML is gratefully acknowledged. 
Critical remarks by Dr. M.~Kun and the referee led to a considerable 
improvement in the presentation of the results.
During this study we made use of the
OGLE data base (http://ogledb.astrouw.edu.pl/$\sim$ogle/CVS/) 
and the McMaster Cepheid Photometry and Radial Velocity Data Archive 
(http://crocus.physics.mcmaster.ca/Cepheid/).

\bsp

\label{lastpage}

\end{document}